\documentclass[twocolumn,prb,showpacs,floatfix]{revtex4}

\usepackage{graphicx}
\usepackage{bm}
\usepackage{color}
\usepackage{amsmath}
\usepackage{amsfonts}
\usepackage{amssymb}

\begin{document}

\preprint{APS/123-QED}

\title{Phase Diagram of Pressure-Induced Superconductivity in EuFe$_2$As$_2$\\
Probed by High-Pressure Resistivity up to 3.2\,GPa}

\author{Nobuyuki Kurita$^{1,2}$, Motoi Kimata$^{1,2}$, Kota Kodama$^{1,3}$, Atsushi Harada$^1$, Megumi Tomita$^1$, Hiroyuki S. Suzuki$^1$, Takehiko Matsumoto$^1$, Keizo Murata$^4$, Shinya Uji$^{1,2,3}$, and Taichi Terashima$^{1,2}$}

\affiliation{$^1$National Institute for Materials Science, Tsukuba, Ibaraki 305-0003, Japan \\
$^2$JST, Transformative Research-Project on Iron Pnictides (TRIP), Chiyoda, Tokyo 102-0075, Japan \\
$^3$Graduate School of Pure and Applied Sciences, University of Tsukuba, Ibaraki 305-0003, Japan \\
$^4$Division of Molecular Materials Science, Graduate School of Science, Osaka City University, Osaka 558-8585, Japan}

\date{\today}

\begin{abstract}
We have constructed a pressure$-$temperature ($P$$-$$T$) phase diagram of $P$-induced superconductivity in EuFe$_2$As$_2$ single crystals, via resistivity ($\rho$) measurements up to 3.2\,GPa. 
As hydrostatic pressure is applied, the temperature  $T_\mathrm{0}$ where an antiferromagnetic (AF) transition of the Fe moments and a structural phase transition occur shifts to lower temperatures, and the corresponding resistive anomaly becomes undetectable for $P$\,$\ge$\,2.5\,GPa.
This suggests that the critical pressure $P_\mathrm{c}$ where $T_\mathrm{0}$ becomes zero is about 2.5\,GPa.
We have found that the AF order of the Eu$^{2+}$ moments survives up to 3.2\,GPa, the highest pressure in the experiments, without significant changes in the AF ordering temperature $T_\mathrm{N}$.
The superconducting (SC) ground state with a sharp transition to zero resistivity at $T_\mathrm{c}$\,$\sim$\,30\,K, indicative of bulk superconductivity, emerges in a pressure range from $P_\mathrm{c}$\,$\sim$\,2.5\,GPa to $\sim$\,3.0\,GPa.
At pressures close to but outside the SC phase, the $\rho(T)$ curve shows a partial SC transition (i.e., zero resistivity is not attained) followed by a reentrant-like hump at approximately $T_\mathrm{N}$ with decreasing temperature.
When nonhydrostatic pressure with a uniaxial-like strain component is applied using a solid pressure medium, the partial superconductivity is continuously observed in a wide pressure range from 1.1\,GPa to 3.2\,GPa.

\end{abstract}

\pacs{74.62.Fj,74.25.Dw,74.25.F-,74.70.Xa}

\maketitle
\section{Introduction}

Over the last decades, understanding of unconventional superconductivity in strongly correlated systems such as cuprate and heavy fermion materials has been one of the most intriguing issues in condensed matter physics.
Despite intensive research efforts, many questions are left unanswered, especially concerning the interplay between superconductivity and magnetism.
The recent discovery of a new class of superconductivity in iron-based materials\,\cite{Kamihara} has opened another route to better understanding the interplay, as well as the high transition temperature ($T_\mathrm{c}$) mechanism.\cite{review}
Shortly after the discovery of superconductivity in LaFeAs(O,F) at $T_\mathrm{c}$\,=\,26\,K,\cite{Kamihara} $T_\mathrm{c}$ was markedly enhanced up to 54\,-\,56\,K\,\cite{Kito2008, ZARen2008a,Wang_56K} in the ``1111" system ($R$FeAsO; $R$\,=\,rare earth).
Thus far, a variety of related compounds with stacked iron-pnictide (or iron-chalcogenide) layers have been found.
The crystal structure spans rather three-dimensional FeSe\,\cite{FeSe} with the simplest form, to highly two-dimensional Sr$_2$ScFePO$_3$\,\cite{perovskite} containing perovskite layers.

EuFe$_2$As$_{2}$, categorized as a moderate two-dimensional ``122" system ($A$Fe$_2$As$_{2}$; $A$\,=\,alkaline earth or Eu) with a tetragonal ThCr$_2$Si$_2$ structure, turns out to be a unique pressure- ($P$-) induced antiferromagnetic (AF) superconductor.\cite{Miclea,Terashima_Eu1,Terashima_Eu2,Kurita_Eu1,Kurita_Eu2,Matsubayashi_Eu}
At ambient pressure, EuFe$_2$As$_{2}$ undergoes an AF order related to the FeAs layers at $T_\mathrm{0}$\,$\sim$\,190\,K,\cite{Raffius_Mossbauer,Ren_poly,Jeevan_single,Xiao_neutron} which is accompanied by a structural transition.\cite{Tegel_Xray}
As commonly reported in other ``1111" and ``122" systems,\cite{Kamihara,Rotter2008b,Sasmal2008, Jiang_BaFe2AsP2,Alireza,Matsubayashi_SrBa} doping\,\cite{Jeevan_EuKFe2As2,Ren_EuFe2AsP2,Jiang_EuFeCo2As2,Zheng_EuSrFeCo2As2} or application of pressure\,\cite{Terashima_Eu1,Matsubayashi_Eu} suppresses the AF/structural transition, and eventually induces bulk superconductivity with the $T_\mathrm{c}$ value of 20\,-\,30\,K.
One distinctive characteristic of EuFe$_2$As$_{2}$ is that the localized Eu$^{2+}$ moments show an AF order at $T_\mathrm{N}$\,$\sim$\,20\,K.\cite{Ren_poly,Jeevan_single,Jiang_NJP09,Xiao_neutron}
The AF order of the Eu$^{2+}$ moments is less sensitive to applied pressure,\cite{Miclea,Terashima_Eu1,Terashima_Eu2,Kurita_Eu1,Kurita_Eu2,Matsubayashi_Eu} and is detectable even inside the bulk superconducting (SC) phase.\cite{Terashima_Eu1,Matsubayashi_Eu}
This leads to the peculiar $T$-dependence of the upper critical field for the $P$-induced superconductivity, as presented in our previous reports.\cite{Terashima_Eu1,Kurita_Eu1}
It has also been suggested from recent high-pressure magnetic and calorimetric measurements\,\cite{Matsubayashi_Eu} that the AF order (Eu$^{2+}$) survives under applied pressure up to about 8\,GPa, above which it probably turns into a ferromagnetic order.

Application of pressure has been established as an excellent probe for precisely tuning ground states, without a random potential or disorder, generally produced by elemental substitutions.
However, particular attention should be paid to the fact that the hydrostaticity of applied pressure is crucially important to investigate the intrinsic superconductivity in iron-based systems.\cite{Yu_Helium,Lee_Ca,Matsubayashi_SrBa,Kotegawa_Pmedium,Yamazaki_Ba} 
On this point, EuFe$_2$As$_{2}$ has a significant advantage because the critical pressure $P_\mathrm{c}$, where $T_\mathrm{0}$\,$\rightarrow$\,0 and the bulk superconductivity appears, is as relatively small as 2.5\,-\,2.7\,GPa,\cite{Miclea,Terashima_Eu1,Terashima_Eu2,Kurita_Eu1,Kurita_Eu2,Matsubayashi_Eu} which can be achieved using a piston-cylinder-type pressure device.
For comparison, the $P_\mathrm{c}$ values for SrFe$_2$As$_{2}$ and BaFe$_2$As$_{2}$ are reported to be 4\,-\,5\,GPa\,\cite{Kotegawa_Pmedium,Matsubayashi_SrBa} and $\sim$\,10\,GPa,\cite{Yamazaki_Ba} respectively.
Thus, EuFe$_2$As$_{2}$ provides a significant opportunity for precisely probing the $P$-induced superconductivity and its interplay with the two different types of magnetism.

In this paper, we present the detailed $P$-$T$ phase diagram of $P$-induced superconductivity in EuFe$_2$As$_{2}$ single crystals, deduced from the high-pressure resistivity measurements up to 3.2\,GPa. 
It is found that, when highly hydrostatic pressure is applied, the $P$-induced SC state emerges with $T_\mathrm{c}$\,$\sim$\,30\,K in a narrow pressure range of 2.5\,GPa to $\sim$\,3.0\,GPa, coexisting with the AF order of Eu$^{2+}$ moments below $T_\mathrm{N}$\,$\sim$\,20\,K.
We also discuss differences in hydrostatic and nonhydrostatic pressure effects on the $P$-induced superconductivity in EuFe$_2$As$_{2}$.

\section{Experimental Details}

\begin{figure}
\begin{center}
\includegraphics[width=0.8\linewidth]{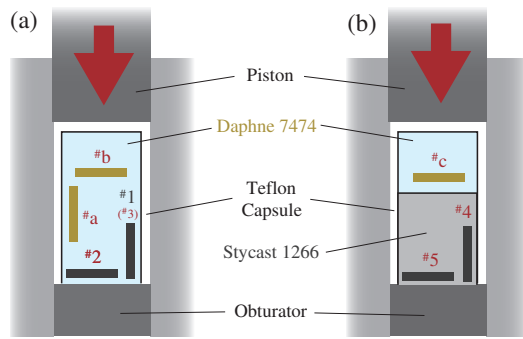}
\end{center}
\caption{(Color online) Simplified schematic views of assemblies for the high-pressure resistivity measurements: (a) samples\,$^{\#}1$ ($^{\#}3$) and $^{\#}2$, and Manganin\,$^{\#}a$ and $^{\#}b$, (b) samples\,$^{\#}4$ and $^{\#}5$, and Manganin\,$^{\#}c$. 
Samples\,$^{\#}1$-$^{\#}3$ and all Manganin wires are immersed in Daphne\,7474 pressure medium, whereas samples\,$^{\#}4$ and $^{\#}5$ are fixed by cured Stycast\,1266.
An external load is applied parallel to the longitudinal axis of a pressure cell as indicated by arrows.
} \label{fig1}
\end{figure}

Single crystals of EuFe$_2$As$_{2}$ were grown by the Bridgman method from a stoichiometric mixture of the constituent elements. 
In this study, we used several crystals labeled as samples\,$^{\#}$1\,-\,$^{\#}$5, which were taken from the same batch (Residual resistivity ratio $RRR$\,=\,7) as used in our previous works.\cite{Terashima_Eu1,Terashima_Eu2,Kurita_Eu1,Kurita_Eu2}
High-pressure resistivity measurements up to 3.2\,GPa have been performed under zero applied magnetic field in a $^4$He cryostat down to 1.6\,K, using a clamped piston cylinder pressure device.\cite{PistonCell}
Electrical resistivity was measured by a conventional four-probe method with an ac current of $I$\,$\sim$\,0.3\,mA (frequency: 10\,$\sim$\,20\,Hz) for the direction $I$\,$\parallel$\,$ab$. 
We used two types of pressure-transmitting medium, Daphne\,7474 (Idemitsu Kosan)\,\cite{Daphne7474} and Stycast\,1266 (cured by Catalyst), to produce a hydrostatic or nonhydrostatic pressure, respectively, as shown in Fig.~\ref{fig1}.
Daphne\,7474 remains in the liquid state up to 3.7\,GPa at room temperature (RT).\cite{Daphne7474}
To avoid an abrupt solidification of the pressure medium, which could be a cause of nonhydrostatic pressure, the cooling speed was regulated at an average value of 0.5\,K/min. 
The generated pressure was determined at 4.2\,K from the relative resistance change of Manganin wires.
The insulation-coated Manganin wires with a diameter of 0.1\,mm and a resistance of 0.7\,$\Omega$/cm, which were coiled a few turns, were exposed to high pressure and low temperature a few times in advance before the measurements.
It should be noted that the resistance of Manganin wires was always measured under the same hydrostatic condition using Daphne\,7474 pressure medium (Fig.~\ref{fig1}).
This permits a direct comparison of the pressure values for all the setups, irrespective of the employed pressure medium, as in Fig.~\ref{fig6}(e).
In this paper, we present three independent resistivity experiments using (i) samples\,$^{\#}$1 and $^{\#}$2 and Manganin\,$^{\#}$a and $^{\#}$b (Daphne\,7474), (ii) sample\,$^{\#}$3 (Daphne\,7474), and (iii) sample\,$^{\#}$4 and $^{\#}$5 (Stycast\,1266).
The assemblies of the samples and the Manganin wires are illustrated in Fig.~\ref{fig1}.
Note that  $P$\,=\,0\,GPa data were obtained by applying a pressure of 0.2\,GPa at RT, considering a reduction of pressure with decreasing temperature.

\section{Results and Discussions}

\subsection{High-Pressure Resistivity Data}

\begin{figure}
\begin{center}
\includegraphics[width=0.95\linewidth]{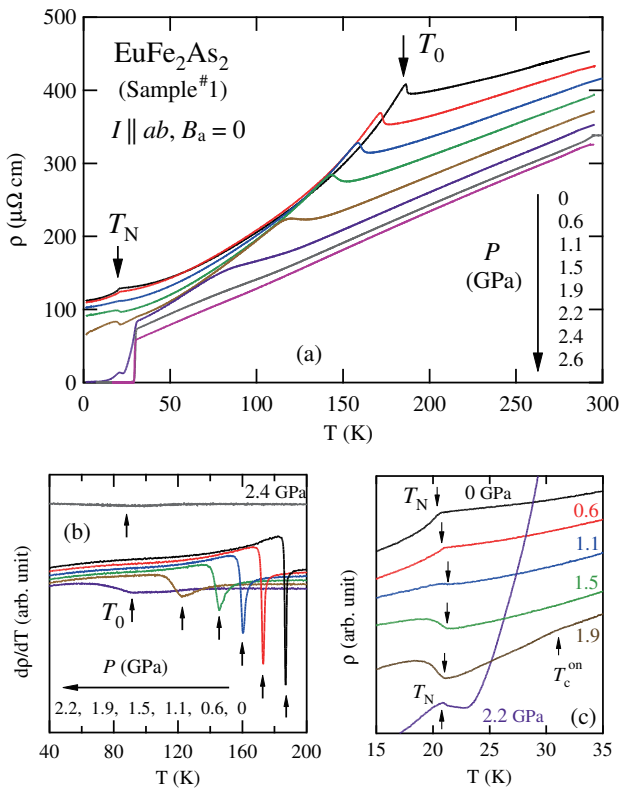}
\end{center}
\caption{(Color online) (a) $\rho$ vs $T$ of EuFe$_2$As$_2$ (sample\,$^{\#}$1; Daphne\,7474) under several pressures up to 2.6\,GPa.
The anomaly at $T_\mathrm{0}$ is due to an AF transition of the Fe moments and a structural phase transition. 
$T_\mathrm{N}$ is an AF ordering temperature of the Eu$^{2+}$ moments.  
(b) d$\rho$/d$T$ vs $T$ for $P$\,$\le$\,2.4\,GPa.
Arrows indicate $T_\mathrm{0}$ where d$\rho$/d$T$ exhibits its minimum. 
(c) Expanded view of $\rho$ vs $T$.
$T_\mathrm{c}^\mathrm{on}$ at 1.9\,GPa indicates an onset temperature of the SC transition.
The data in (b) and (c) are vertically shifted for clarity.
} \label{fig2}
\end{figure}

\begin{figure}
\begin{center}
\includegraphics[width=0.95\linewidth]{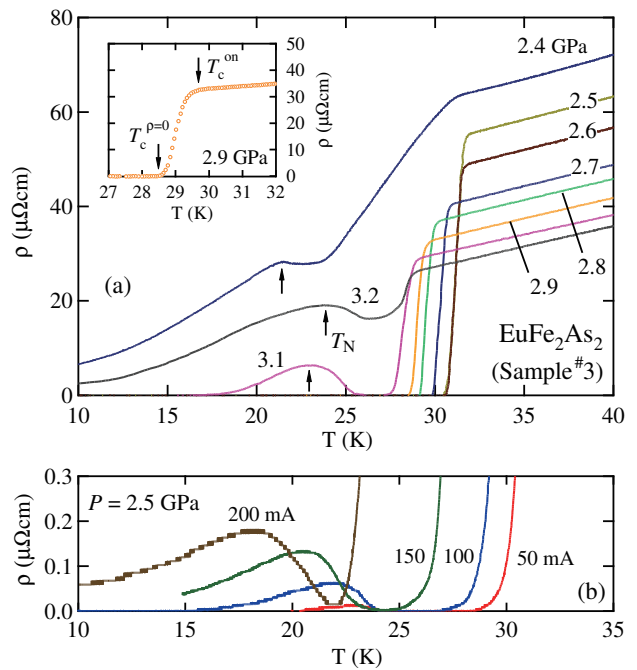}
\end{center}
\caption{(Color online) (a) Low-temperature expanded view of $\rho(T)$ data for EuFe$_2$As$_2$ (sample\,$^{\#}$3; Daphne\,7474) under several pressures up to 3.2\,GPa. 
As illustrated in the inset using the data at 2.9\,GPa, $T_\mathrm{c}^\mathrm{on}$ and $T_\mathrm{c}^{\rho=0}$ are determined by the onset and  $\rho$\,=\,0 of the SC transition.
(b) $\rho(T)$ data at 2.5\,GPa for several current values up to $I$\,=\,200\,mA ($\sim$\,500\,A/cm$^2$).
Shifts of SC transitions and resistive humps to lower temperatures as current increases may not be intrinsic, and are most likely attributed to a Joule heating effect. } \label{fig3}
\end{figure}

Figure~\ref{fig2}(a) shows the temperature dependence of the resistivity, $\rho(T)$, of EuFe$_2$As$_2$ (sample\,$^{\#}$1; Daphne\,7474) under hydrostatic pressures up to 2.6\,GPa.
At 0\,GPa, $\rho(T)$ shows cusp-like behavior at $T_\mathrm{0}$\,=\,188\,K as indicated by an arrow.
This is attributed to the AF transition related to the FeAs layers,\cite{Raffius_Mossbauer,Ren_poly,Jeevan_single,Xiao_neutron} which coincides with a structural transition from tetragonal to orthorhombic symmetry.\cite{Tegel_Xray}
With increasing pressure, the resistive cusp is gradually suppressed and shifts to lower temperatures.
As indicated by arrows in Fig.~\ref{fig2}(b), we define $T_\mathrm{0}$ as the temperature where d$\rho$/d$T$ exhibits its minimum. 
The minimum in d$\rho$/d$T$ can be identified up to 2.4\,GPa, although only faintly visible at 2.4\,GPa.
At 2.6\,GPa, resistivity follows approximately $T$-linear behavior in a broad temperature range above $T_\mathrm{c}$\,=\,30\,K without a noticeable signature associated with $T_\mathrm{0}$, consistent with previous reports.\cite{Terashima_Eu1,Kurita_Eu1,Kurita_Eu2}
Thus, the critical pressure $P_\mathrm{c}$, where $T_\mathrm{0}$\,$\rightarrow$\,0, is about 2.5\,GPa.\cite{Pc}
It remains to be resolved whether the simultaneous AF/structural transition in EuFe$_2$As$_2$ splits into two individual transitions by applying pressure.
A separation of the two transitions might explain broad transitions observed under high pressure.
The separation has been observed in ``1111"\,\cite{T0_LaFeAsO} and doped ``122" compounds\,\cite{T0_BaKFe2As2} at ambient pressure.

As shown by arrows in Fig.~\ref{fig2}(c), the AF order of the localized Eu$^{2+}$ moments under applied pressure is evidenced by distinctive changes in $\rho(T)$ around 20\,K.
It is of interest that the shape of the $\rho(T)$ curve near $T_\mathrm{N}$ varies with increasing pressure.
For low pressures ($P$\,$\le$\,0.6\,GPa), $\rho(T)$ exhibits a sudden decrease below $T_\mathrm{N}$.
For 1.1\,$\le$\,$P$\,$\le$\,1.9\,GPa, $\rho(T)$ shows an enhancement at $T_\mathrm{N}$ followed by a peaklike behavior as temperature decreases.
A similar feature in $\rho(T)$ at $T_\mathrm{N}$ has also been observed in several AF metals with localized moments and is generally attributed to a superzone effect arising from a gap formation associated with a periodic arrangement of the magnetic moments.\cite{superzone}
At 1.9\,GPa, one can see a resistive kink at 31\,K, which we ascribe as an onset temperature of a SC transition ($T_\mathrm{c}^\mathrm{on}$).
With increasing pressure, the kink evolves into a marked decrease in resistivity below $T_\mathrm{c}^\mathrm{on}$, and the SC transition attains zero resistivity at 2.4\,GPa for this sample, although the transition width is very broad.
At 2.6\,GPa ($>$\,$P_\mathrm{c}$), where the bulk superconductivity was confirmed by our ac-susceptibility ($\chi$) measurements,\cite{Terashima_Eu1} $\rho(T)$ shows a sharp SC transition to zero resistivity.
The humplike behavior attributed to $T_\mathrm{N}$ is distinctly signaled below $T_\mathrm{c}^\mathrm{on}$ for 1.9\,GPa and 2.2\,GPa, but not for 2.4\,GPa and 2.6\,GPa.

Figure~\ref{fig3}(a) shows the low-temperature magnification of $\rho(T)$ data for EuFe$_2$As$_2$ (sample\,$^{\#}$3; Daphne\,7474), under several pressures up to 3.2\,GPa.
Unlike the result for the sample\,$^{\#}$1, the $\rho(T)$ curve at 2.4\,GPa shows a partial SC transition, but it does not reach zero resistivity.
As pressure increases, the SC transition gradually shifts to lower temperatures. 
At 3.1\,GPa, with decreasing temperature, $\rho(T)$ exhibits a slightly broader transition to zero resistivity, followed by a humplike behavior, and then returning to zero resistivity again.
At 3.2\,GPa, the $\rho(T)$ curve with the humplike behavior no longer reaches zero down to the lowest temperature of 1.6\,K.
It is noted that the resistive hump is observed at similar temperatures at 2.4\,GPa and 3.2\,GPa.
This is consistent with the results of $\rho$, ac-$\chi$ and ac-calorimetry measurements under higher pressure,\cite{Matsubayashi_Eu} which demonstrate that the AF order of the Eu$^{2+}$ moments remains up to $\sim$\,8\,GPa with only a moderate increase in $T_\mathrm{N}$.

The width of the SC transition $\Delta T_\mathrm{c}$, defined as $T_\mathrm{c}^\mathrm{on}$\,$-$\,$T_\mathrm{c}^{\rho=0}$, is estimated to be 1.5\,K at 2.5\,GPa [see the inset of Fig.~\ref{fig3}(a) for the definition of $T_\mathrm{c}^\mathrm{on}$ and $T_\mathrm{c}^\mathrm{\rho=0}$].
As shown in Fig.~\ref{fig4}, with increasing pressure, the transition becomes sharper, and, at 2.7\,GPa, $\Delta T_\mathrm{c}$ shows a minimum value of 1\,K (or 0.6\,K when the width is defined by the 90\,\% and 10\,\% values of the normal state resistivity at $T_\mathrm{c}^\mathrm{on}$, as often used in other reports).

Figure~\ref{fig3}(b) shows $\rho(T)$ data of EuFe$_2$As$_2$ at 2.5\,GPa ($\sim$\,$P_\mathrm{c}$) under several current values up to $I$\,=\,200\,mA, which corresponds to the current density of $\sim$\,500\,A/cm$^2$.
At 0.3\,mA, the $\rho(T)$ curve exhibits a sharp SC transition without a reentrant-like hump due to the AF order of the Eu$^{2+}$ moments.
At 50\,mA, one can slightly see an appearance of the humplike behavior around 22\,K, most likely attributed to $T_\mathrm{N}$.
With increasing current up to 200\,mA, the hump becomes more noticeable, although the resistivity peak is only 0.2\,$\mathrm{\mu \Omega}$\,cm at 200\,mA, $\sim$\,0.4\,\% of the normal state resistivity at $T_\mathrm{c}^\mathrm{on}$.
The appearance of the hump probably means that the magnitude of the critical current $J_\mathrm{c}$ is reduced around $T_\mathrm{N}$ with the development of the AF order of the Eu$^{2+}$ moments.
A similar reentrant-like hump near $T_\mathrm{N}$ was reported in several AF superconductors such as NdRhB$_4$,\cite{NdRh4B4} Gd$_{1.2}$Ho$_6$S$_8$,\cite{GdMo6S8} and HoNi$_2$B$_2$C.\cite{HoNi2B2C}
It must be noted that a Joule heating effect is the main reason why the SC transition, as well as the resistive hump, has shifted to lower temperatures with an increase in the current.\cite{IVlinearlity}
However, this does not affect the above discussion.

\subsection{Pressure-Temperature Phase Diagram}

\begin{figure}
\begin{center}
\includegraphics[width=0.95\linewidth]{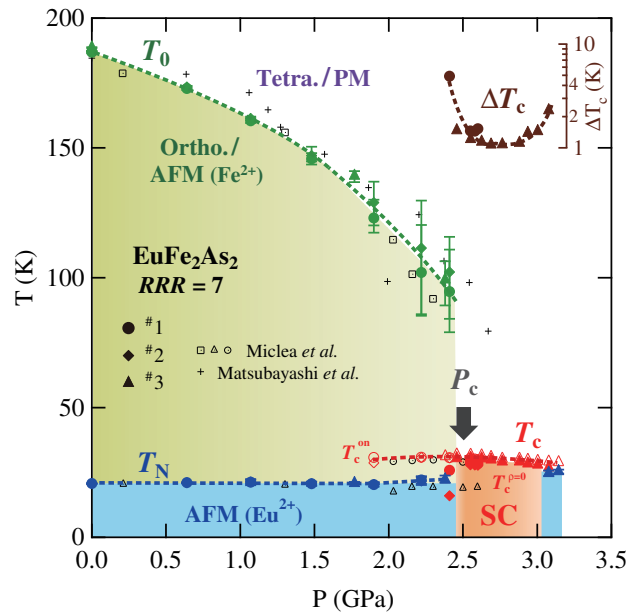}
\end{center}
\caption{(Color online) $P$\,$-$\,$T$ phase diagram of $T_\mathrm{0}$, $T_\mathrm{N}$, $T_\mathrm{c}$, and $\Delta T_\mathrm{c}$ (=\,$T_\mathrm{c}^\mathrm{on}$\,$-$\,$T_\mathrm{c}^{\rho=0}$) in EuFe$_2$As$_2$ (samples\,$^{\#}$1\,-\,$^{\#}$3; Daphne\,7474), deduced from the $\rho(T)$ data up to 3.2\,GPa.
PM, AFM and SC indicate the paramagnetic, antiferromagnetic and superconducting states, respectively. 
For the SC state, open and solid symbols indicate $T_\mathrm{c}^\mathrm{on}$ and $T_\mathrm{c}^{\rho=0}$, respectively. 
As indicated by an arrow, the critical pressure ($P_\mathrm{c}$), where $T_\mathrm{0}$\,$\rightarrow$\,0, is about 2.5\,GPa.
Dashed curves are eye guides.
For comparison, data from Refs.\,\onlinecite{Miclea} and \onlinecite{Matsubayashi_Eu} are also shown.} \label{fig4}
\end{figure}

In Fig.~\ref{fig4}, we illustrate the $P$$-$$T$ phase diagram of the characteristic temperatures, $T_\mathrm{0}$, $T_\mathrm{N}$, and $T_\mathrm{c}$ ($T_\mathrm{c}^\mathrm{on}$ and $T_\mathrm{c}^{\rho=0}$) in EuFe$_2$As$_2$ (samples\,$^{\#}$1\,-\,$^{\#}$3; Daphne\,7474), determined from the $\rho(T)$ data under hydrostatic pressures up to 3.2\,GPa.
Our findings on the phase diagram are summarized as (i)-(iv).
(i) The pressure evolution of $T_\mathrm{0}$ and $T_\mathrm{N}$ is consistent with that obtained by Miclea {\it et al.} (Ref.\onlinecite{Miclea}) and Matsubayashi {\it et al.} (Ref.\onlinecite{Matsubayashi_Eu}).
$T_\mathrm{0}$ disappears for $P$\,$\ge$\,2.5\,GPa ($P_\mathrm{c}$\,$\sim$\,2.5\,GPa), whereas $T_\mathrm{N}$ survives up to at least 3.2\,GPa.
(ii) $T_\mathrm{c}^\mathrm{on}$ emerges from 1.9\,GPa, although the partial SC transition with nonzero resistivity is probably filamentary for $P$\,$<$\,2.4\,GPa, as was revealed by previous ac-$\chi$ measurements at 2.2\,GPa.\cite{Terashima_Eu1}
Likewise, nonbulk superconductivity for the 3.2\,GPa data can be inferred from the similar partial SC transition.
(iii) At 2.4\,GPa ($\lesssim$\,$P_\mathrm{c}$), the SC transition attains zero resistivity for some samples such as samples\,$^{\#}$1 and $^{\#}$2 (not for sample\,$^{\#}$3), but the width is broadened ($\Delta T_\mathrm{c}$\,$\sim$\,5\,K).
This is likely due to an internal strain and/or composition inhomogeneity, which may bring $T_\mathrm{0}$\,$\rightarrow$\,0 at some parts of a crystal, and consequently induce partial superconductivity.
A similar scenario can also be applied to the 3.1\,GPa data where the width of the SC transition to zero resistivity is broadened.
(iv) Bulk superconductivity exists in narrow pressures of 2.5\,GPa ($\sim$\,$P_\mathrm{c}$) to $\sim$\,3.0\,GPa, judging from the sharp SC transitions to zero resistivity.

Ref.\onlinecite{Miclea} shows that a partial SC transition with a reentrant-like hump is continuously observed up to 2.6\,GPa even above $P_\mathrm{c}$, whereas it is noticeable only at the outer boundaries of the SC phase ($P$\,$<$\,$P_\mathrm{c}$ or $P$\,$>$\,3\,GPa) in our case.
As claimed in Ref.\,\onlinecite{Matsubayashi_Eu} and supported by our results in the following section, the absence of zero-resistivity transition in Ref.\,\onlinecite{Miclea} may be caused by a nonhydrostatic pressure effect.

The loss of bulk superconductivity above $\sim$\,3\,GPa, which we found, is in conflict with a previous claim in Ref.\,\onlinecite{Uhoya_Eu} that a $P$-induced SC state in EuFe$_2$As$_2$ exists up to $\sim$\,16\,GPa, and that the $T_\mathrm{c}$ value attains its maximum of 41\,K around 10\,GPa.
In Ref.\,\onlinecite{Uhoya_Eu}, tiny resistivity drops observed in the wide pressure range were attributed to SC transitions.
However, those drops are most likely due to magnetic ordering of the Eu$^{2+}$ moments as clearly demonstrated by magnetic measurements on EuFe$_2$As$_2$ in Ref.\,\onlinecite{Matsubayashi_Eu}.
The claim in Ref.\,\onlinecite{Guo_Eu} that EuFe$_2$As$_{1.4}$P$_{0.6}$ superconducts up to $\sim$\,20\,GPa is also questionable, since the superconductivity is based only on observations of slight resistivity drops.

Recently, systematic studies of P-doping effects at As-site in EuFe$_2$As$_2$ have been reported.\cite{Jeevan_chemicalP}
The phase diagram as a function of dopant concentration is roughly consistent with that as a function of pressure.
For both cases, the SC phases are confined in a narrow dopant/pressure regime, whereas magnetic orders of the Eu$^{2+}$ moments remain in all dopant levels or up to extremely high pressures ($\sim$\,20\,GPa\,\cite{Matsubayashi_Eu}).
On the other hand, there seems to be a difference between the two tuning parameters regarding at which concentration/pressure the magnetic order of the Eu$^{2+}$ moments changes from AF to FM.\cite{Matsubayashi_Eu,Jeevan_chemicalP}
The AF/FM transition occurs at about 8\,GPa, a much higher pressure than that of the SC phase for the pressure case,\cite{Matsubayashi_Eu} whereas it coincides with the boundary of the SC phase on the overdoping side for the doping case.\cite{Jeevan_chemicalP} 

\subsection{Hydrostatic vs Nonhydrostatic Pressure}

\begin{figure}
\begin{center}
\includegraphics[width=0.95\linewidth]{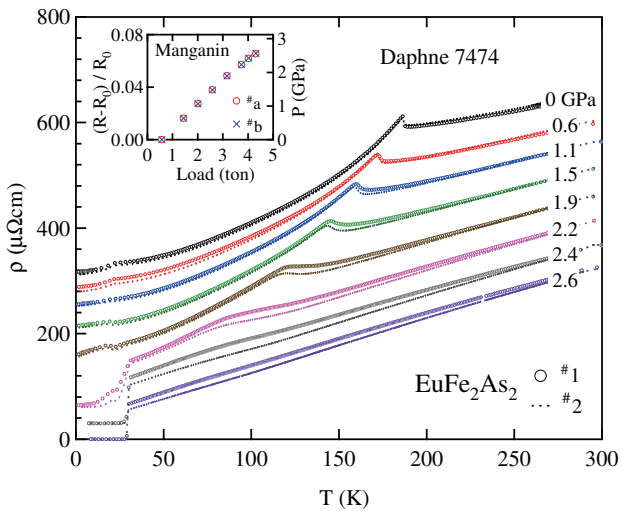}
\end{center}
\caption{(Color online) Comparison of $\rho(T)$ data of EuFe$_2$As$_2$ for samples $^{\#}1$ (open circles) and $^{\#}2$ (dotted curves),   simultaneously measured using Daphne\,7474 pressure medium.
The data are arbitrarily shifted in the longitudinal direction for clarity.
The inset displays a resistance change ratio $\Delta R$ (left axis) and the corresponding pressure value (right axis) obtained at 4.2\,K for Manganin\,$^{\#}a$ and $^{\#}b$ as a function of the externally applied load.
Sample\,$^{\#}1$ and Manganin\,$^{\#}a$ are aligned parallel to the load axis, and sample\,$^{\#}2$ and Manganin\,$^{\#}b$ are in the perpendicular direction, as shown in Fig.~\ref{fig1}(a).
} \label{fig5}
\end{figure}

\begin{figure}
\begin{center}
\includegraphics[width=0.95\linewidth]{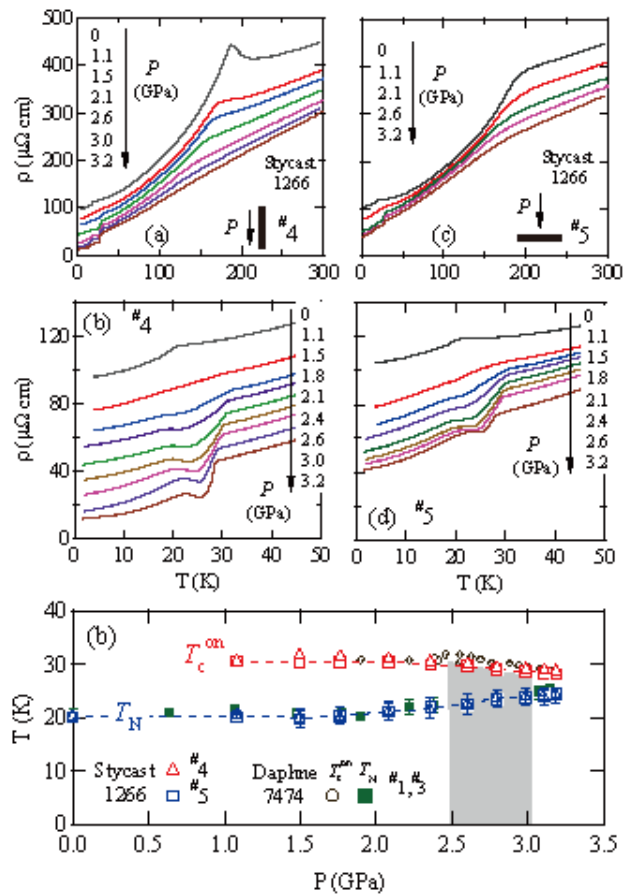}
\end{center}
\caption{(Color online) $\rho$ vs $T$ of EuFe$_2$As$_2$ at several pressures up to 3.2\,GPa for samples $^{\#}4$ (a) and $^{\#}5$ (b), simultaneously measured using a Stycast\,1266 pressure medium.
Samples\,$^{\#}4$ and $^{\#}5$ are aligned parallel and perpendicular, respectively, to the load direction as in Fig.~\ref{fig1}(b).
(c) and (d) show the low-temperature expanded views for samples\,$^{\#}4$ and $^{\#}5$, respectively. 
(e) Pressure evolutions of $T_\mathrm{N}$ and $T_\mathrm{c}^\mathrm{on}$ for samples\,$^{\#}4$ ($\triangle$) and $^{\#}5$ ($\square$).
For comparison, $T_\mathrm{N}$ ($\blacksquare$) and $T_\mathrm{c}^\mathrm{on}$ ($\bigcirc$) obtained using Daphne\,7474 pressure medium for the samples\,$^{\#}1$ and $^{\#}3$ are also shown.
A shaded area represents a SC phase where the resistivity is zero under hydrostatic pressure.} \label{fig6}
\end{figure}	

Although Daphne\,7474 remains in a liquid state up to 3.7\,GPa at RT,\cite{Daphne7474} it eventually solidifies with decreasing temperature.
We have therefore checked the hydrostaticity of the generated pressure at low temperatures by performing simultaneous resistivity measurements using two Manganin wires and two EuFe$_2$As$_2$ crystals: Manganin\,$^{\#}a$ and sample\,$^{\#}1$ are mounted parallel to the load axis, and Manganin\,$^{\#}b$ and sample\,$^{\#}2$ in the perpendicular direction, as displayed in Fig.~\ref{fig1}(a).
Since a Manganin wire is highly sensitive to nonhydrostatic pressure, two sets of Manganin wires aligned as Manganin\,$^{\#}a$ and $^{\#}b$ are often used to estimate the degree of hydrostaticity of a pressure medium.\cite{Eremets}
When the pressure medium becomes nonhydrostatic, the resistance change ratio $\Delta R$ [=\,($R$\,$-$\,$R_0$)/$R_0$ ($R_0$: resistance at ambient pressure)] is expected to deviate significantly from each other.
In the case of Daphne\,7474, as seen in the upper inset of Fig.~\ref{fig5}, $\Delta R$ for Manganin\,$^{\#}a$ and $^{\#}b$ estimated at 4.2\,K is almost identical up to the investigated highest pressure of 2.6\,GPa ($>$\,$P_\mathrm{c}$).
This indicates that the low-temperature pressure produced via the Daphne\,7474 pressure medium is satisfactorily hydrostatic up to  at least 2.6\,GPa, where the $P$-induced superconductivity shows up in EuFe$_2$As$_2$.

The main panel of Fig.~\ref{fig5} shows $\rho(T)$ data of EuFe$_2$As$_2$ for samples\,$^{\#}1$ (circle symbols) and $^{\#}2$ (dotted curves) up to 2.6\,GPa.
For clarity, the data are arbitrarily shifted in the longitudinal direction.
The pressure evolutions of the two $\rho(T)$ curves are similar to each other, as can also be inferred from the phase diagram in Fig.~\ref{fig4}, suggesting no significant nonhydrostatic pressure effect.
The small difference between the data is likely attributed to the sample dependence.

Finally, we discuss the effects of nonhydrostatic pressures with uniaxial-like strain components on the SC phase diagram of EuFe$_2$As$_2$.
Figure~\ref{fig6} shows $\rho(T)$ data of EuFe$_2$As$_2$ at several pressures up to 3.2\,GPa for the samples (a) $^{\#}4$ and (b) $^{\#}5$, simultaneously measured using Stycast\,1266 (cured by Catalyst) as a solid pressure medium.
Samples\,$^{\#}4$ and $^{\#}5$ are mounted parallel and perpendicular, respectively, to the load axis [see Fig.~\ref{fig1}(b)].
In this setup, samples\,$^{\#}4$ and $^{\#}5$ are expected to receive a uniaxial-like strain more on the $ab$-plane and $c$-axis directions, respectively, as in the illustrations.\cite{uniaxialP}
Interestingly, application of 0.2\,GPa at RT ($P$\,=\,0\,GPa data) produces contrasting $\rho(T)$ curves between the two different alignments, although $\rho(T)$ curves at higher pressures are qualitatively similar to each other.
For both cases, the resistive transition associated with $T_\mathrm{0}$ is rapidly broadened with increasing pressure, which makes it difficult to unambiguously determine $T_\mathrm{0}$ values.
However, it is noted that a convex curvature associated with $T_\mathrm{0}$ is still observable at 3.2\,GPa for both samples.

As shown in Figs.~\ref{fig6}(b) and (d), $\rho(T)$ curves for two alignments show partial SC transitions below $T_\mathrm{c}^\mathrm{on}$\,$\sim$\,30\,K with reentrant-like behavior near $T_\mathrm{N}$ but never reach zero resistivity, in a wide pressure range from 1.1\,GPa to 3.2\,GPa.
The result is similar to that obtained using silicon oil\,\cite{Miclea} and Daphne\,7373\,\cite{Matsubayashi_Eu} pressure media.
It is found that the uniaxial-like strain parallel to the $c$-axis is more detrimental to the superconductivity than that parallel to the $ab$-plane, since the drop in resistivity below $T_\mathrm{c}^\mathrm{on}$ is more salient with a much smaller residual resistivity for the latter case.
This may relate to the fact that the convex curvature due to $T_\mathrm{0}$ is more suppressed when the uniaxial-like strain is applied parallel to the $ab$-plane.
Thus, a suppression of $T_\mathrm{0}$ could be an important key to stabilizing the superconductivity.
Indeed, bulk superconductivity with zero resistivity emerges with a complete suppression of $T_\mathrm{0}$ in the case of Daphne\,7474.

Figure~\ref{fig6}(e) illustrates the pressure evolutions of $T_\mathrm{N}$ and $T_\mathrm{c}^\mathrm{on}$ for samples\,$^{\#}4$ and $^{\#}5$ (Stycast\,1266) up to 3.2\,GPa.
For comparison, $T_\mathrm{N}$ and $T_\mathrm{c}^\mathrm{on}$ obtained using Daphne\,7474 pressure medium for the sample\,$^{\#}3$ are also shown.
The shaded area corresponds to the SC phase under hydrostatic pressure where the SC transition attains zero resistivity, indicative of bulk superconductivity.
Notably, $T_\mathrm{N}$ for all the samples exhibits quite similar pressure evolutions with a moderate increase, independently of the pressure medium and of the direction of uniaxial-like strain.
On the other hand, the phase diagram obtained via Stycast\,1266 is crucially different from that via Daphne\,7474, in terms of the superconductivity.
When Stycast\,1266 is used, an onset of the SC transition can be found from a lower pressure of 1.1\,GPa (compared with 1.9\,GPa in the case of Daphne\,7474), and resistivity does not reach zero up to 3.2\,GPa.
These facts indicate that nonhydrostatic pressure can induce a partial SC transition at low pressures but is detrimental to the bulk superconductivity, whereas it has no significant effect on the AF order of the Eu$^{2+}$ moments.  

\section{Summary}

High-pressure resistivity measurements up to 3.2\,GPa have been performed on EuFe$_2$As$_2$ single crystals to establish the phase diagrams of the $P$-induced superconductivity. 
We have found that a sharp SC transition to zero resistivity, indicating bulk superconductivity, appears in a limited pressure range of 2.5\,GPa ($\sim$\,$P_\mathrm{c}$) to $\sim$\,3.0\,GPa, with a complete suppression of $T_\mathrm{0}$ by applying hydrostatic pressure using Daphne\,7474 pressure medium.
By contrast, application of nonhydrostatic pressure produced via Stycast\,1266 brings only a partial SC transition without zero resistivity in a broad pressure range for $P$\,$\ge$\,1.1\,GPa.
It is also found that the AF order of the Eu$^{2+}$ moments persists up to the highest measured pressure of 3.2\,GPa with a moderate increase in $T_\mathrm{N}$, independently of the hydrostaticity of applied pressure.

\section*{Acknowledgment}
We would like to thank A. Mitsuda and K. Matsubayashi for fruitful discussions.


\begin{thebibliography}{25}
\bibitem{Pc}
$P_\mathrm{c}$ varies depending on the quality of the samples.\cite{Kurita_Eu1}
For samples\,$^{\#}$1-$^{\#}$5 from the batch with $RRR$\,=\,7, $P_\mathrm{c}$ is approximately 2.5\,GPa.
\bibitem{IVlinearlity} 
A violation of a current-voltage linearity can be inferred from the fact that, with increasing current density, the normal state resistivity at $T_\mathrm{c}^\mathrm{on}$ remains mostly unchanged in spite of a nonnegligible decrease in $T_\mathrm{c}$.
\bibitem{uniaxialP}
After the measurements, we confirmed that the Stycast\,1266 had a dent, indicating a stress beyond the elastic limit, only to the load direction.
\end{thebibliography}
\end{document}